\documentclass[11pt]{article}
\usepackage{amssymb,amsmath}
\usepackage{doublespace}
\usepackage{overcite}
\usepackage{epsf}

\begin{document}
\begin{spacing}{1.0}
\title{
HARVESTING THERMAL FLUCTUATIONS: ACTIVATION PROCESS INDUCED
BY A NONLINEAR CHAIN IN THERMAL EQUILIBRIUM}

\author{Ramon Reigada\footnote{Permanent address:
Departament de Qu\'{\i}mica-F\'{\i}sica, Universitat de Barcelona,
Avda. Diagonal 647, 08028 Barcelona, Spain}\\
Department of Chemistry and Biochemistry 0340\\
University of California, San Diego\\
La Jolla, California 92093-0340\\
\and
Antonio Sarmiento\footnote{Permanent address:
Instituto de Astronom\'{\i}a, Apdo. Postal 70-264, Ciudad Universitaria,
M\'{e}xico D. F. 04510, M\'{e}xico}\\
Department of Chemistry and Biochemistry 0340\\
University of California, San Diego\\
La Jolla, California 92093-0340\\
\and
Aldo H. Romero\footnote{Present address: Max-Planck Institut f\"{u}r
Festk\"{o}rperforschung, Heisenbergstr. 1, 70569 Stuttgart, Germany}\\
Department of Chemistry and Biochemistry 0340\\
and Department of Physics\\
University of California, San Diego\\
La Jolla, California 92093-0340\\
\and
J. M. Sancho\\
Departament d'Estructura i Constituents de la Mat\`{e}ria\\
Universitat de Barcelona\\
Avda. Diagonal 647, 08028 Barcelona, Spain\\
\and
Katja Lindenberg\\
Department of Chemistry and Biochemistry 0340\\
and Institute for Nonlinear Science\\
University of California San Diego\\
La Jolla, California 92093-0340}
\date{\today }
\maketitle

\end{spacing}
\begin{spacing}{1.5}

\begin{abstract}
We present a model in
which the immediate environment of a bistable system is a
molecular chain which in turn is connected to a thermal
environment of the Langevin form.  The
molecular chain consists of masses connected by harmonic or by
anharmonic springs. The distribution, intensity, and mobility of
thermal fluctuations in these chains is strongly dependent on the
nature of the springs and leads to different transition dynamics
for the activated process.  Thus, all else (temperature, damping,
coupling parameters between the chain and the bistable system)
being the same, the hard chain may provide an environment
described as diffusion-limited and more effective in the
activation process, while the soft chain may provide an
environment described as energy-limited and less effective.  The
importance of a detailed understanding of the thermal environment
toward the understanding of the activation process itself is thus
highlighted.

\end{abstract}

\section{Introduction}
\label{intro
}
The search for mechanisms that may induce the
spontaneous localization of vibrational energy in molecular
materials has surfaced in a variety of contexts where such
localized energy may then trigger other events.  These may include
switching and other threshold phenomena, chemical reactions, local
melting and other deformational effects, and even detonation.
In the Kramers
problem~\cite{Kramers,MelnikovHanggi} a particle moving in
a bistable potential is used as a model for a chemical process.
The trajectory of the particle is associated with the reaction
coordinate (RC). One well of the bistable potential represents the
``reactant" state, the other the ``product" state, and separating
them is the ``activation barrier."  The bistable potential is
connected to a thermal environment, typically through fluctuating
and dissipative terms, and every once in a while a large thermal
fluctuation causes the particle to surmount the barrier and move
from one well to the other.  The average rate of occurrence of
these events is associated with the reaction rate.
This mesoscopic Langevin-type of approach admits of
an underlying microscopic description of the thermal environment
and its coupling to the bistable system.  For instance, the usual
Langevin equation with an instantaneous dissipation and Gaussian
$\delta$-correlated fluctuations can be derived from a
picture in which the system is harmonically coupled to an infinite
number of harmonic oscillators with a uniform spectrum.
A generalized Langevin picture involving
dissipative memory terms and correlated fluctuations
is associated with a more complex spectrum~\cite{Lindenberg}.
It is clear, and has become a topic
of considerable interest, that the nature of the environment and
its coupling to the bistable system profoundly influence
the transition rate.

A different but related set of problems that has attracted
intense interest in recent years concerns the spontaneous
localization of vibrational energy in periodic nonlinear arrays.
The pioneering work of Fermi, Pasta and Ulam~\cite{FPU} demonstrated
that a periodic lattice of coupled nonlinear oscillators is not
ergodic, and that energy in such a lattice may never be distributed
uniformly. A great deal of work has since followed in an attempt
to understand how energy is distributed in discrete nonlinear
systems~\cite{Lepri,Hu,Allen,Bourbonnais,Visco,Willis,Dauxois,Szeftel,Tsironis,Bilbault,Brown,Ebeling1,Takeno}.
The existence of solitons and more generally of breathers
and other energy-focusing mechanisms,
and the stationarity or periodic recurrence or even
slow relaxation of such spatially localized excitations, are viewed
as nonlinear  phenomena with important consequences
in many physical systems~\cite{Tsironis,Zakharov,Bulsara}.
The search for localization mechanisms that are robust  even when
the arrays are in a thermal environment~\cite{Tsironis,Brown,Ebeling1,Reigada1}
has, on
the one hand, narrowed the problem (because some localization
mechanisms are fragile against thermal fluctuations) but on the
other hand broadened it (because new entropy-driven localization
mechanisms become possible).  Thermal effects may be particularly
important in biophysical and biochemical applications
at the molecular level~\cite{Bishop,Gottberg,Wolynes}.

The interest in the distribution and motion of energy in {\em
periodic}
arrays arises in part because localized energy in these systems may
be {\em mobile},
in contrast with systems where energy localization occurs through
disorder. Localized energy that moves with little or no dispersion
may appear at one location on an array and may then
be able to move to another where it can be used in a subsequent
process. 
Traditional harmonic models suffer from the fact that
dispersion thwarts such a mechanism for energy transfer.
There has been a surge of recent activity in an attempt to understand
the thermal conductivity of nonlinear chains~\cite{Prosen,Aoki,Hu2}

\begin{figure}[htb]
\begin{center}
\hspace{3.0in}
\epsfxsize = 4.5in
\epsffile{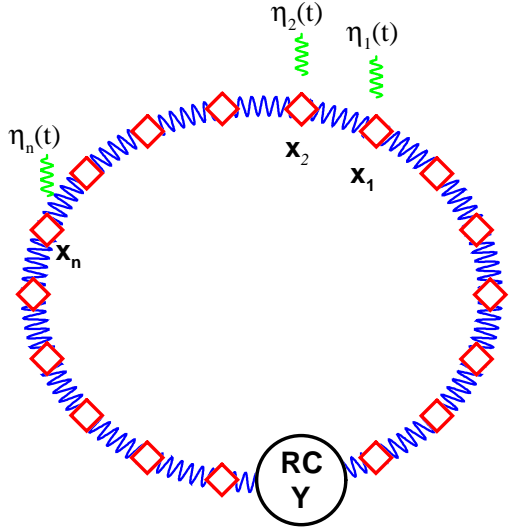}
\vspace{-0.5in}
\end{center}
\caption
{Schematic of a bistable impurity (``reaction coordinate" RC) connected
to a chain that interacts with a heat bath at temperature $T$.  The chain
masses are indicated by rhombuses. Their interactions (shown schematically
as springs) may be harmonic or anharmonic.  Each mass in the chain is
subject to thermal fluctuations denoted by $\eta$ and the usual
accompanying dissipation. The bistable impurity interacts only with the
chain, which thus provides its thermal environment.
The bistable system is inserted in the chain;
its detailed interaction with the chain is discussed in the text.}
\label{drawings}
\end{figure}

The connection between the study of perfect nonlinear arrays and
the Kramers problem
arises because such arrays may themselves serve as models for a heat bath
for other systems connected to them \cite{Sancho,Ebeling2,Casado}.
Albeit in different
contexts, ``perfect" arrays serving as energy storage and transfer
assemblies for chemical or photochemical processes are not
uncommon~\cite{Schulten,Scott,Bolton,Cruzeiro}, and literature on the subject
goes back for two decades~\cite{Collins,Pnevmatikos,StMalo}. 
We thus consider the following variant of the Kramers problem: a
bistable system connected to a nonlinear chain, which is in turn
connected to a heat bath in the usual Langevin manner (see
Fig.~\ref{drawings}).  The
bistable system is only connected to the environment through its
embedding in the nonlinear chain, and therefore the ability of the
chain to spontaneously localize thermal energy and perhaps to
transport it to the location of the bistable system can profoundly
affect the transition rate.  We investigate the behavior of this
model for different types of anharmonic chains and thereby
establish the important role of the nature of the environment on
these chemical model systems.

In Section~\ref{chains} we discuss the energy landscape typical of
various nonlinear chains in thermal equilibrium.
In Section~\ref{kramers} the variant of the
Kramers problem wherein a bistable system is connected to each of
the different chains is presented.  Section~\ref{results} details our
results for the transition statistics in the bistable system.
We compare and contrast the transition statistics in the different
chains and compare them to those found in the standard (Markovian)
Kramers and generalized Kramers problems.  We conclude with a summary
and some notes on future directions in Section~\ref{conclusions}.

\section{Nonlinear Chains}
\label{chains}

The simplest nonlinear periodic arrays consist of masses connected by
springs that may be harmonic or anharmonic.
The masses may also experience a
local harmonic or anharmonic potential.  In a recent paper we
presented a detailed view of the thermal landscape of  arrays
with local hard (the ``$\phi^4$ model"), harmonic, or soft potentials
and harmonic
interactions~\cite{Reigada1}. Here we present
the complementary analysis (more interesting, it turns out, in the
context of the Kramers problem) of the thermal landscape of masses
connected by anharmonic springs (with no local potentials).

The spontaneous localization of
energy in any system in thermal equilibrium
is simply a reflection of the thermal fluctuations described by
statistical mechanics and is unrelated to system dynamics.  On the
other hand,
the way in which these fluctuations dissipate and/or move and
disperse, that is, the temporal evolution of thermal fluctuations,
is dictated by the system dynamics and, in particular, by
the channels connecting the chain to the thermal environment (dissipation)
and the masses to one another (intermolecular interactions).

We pose the following questions: 1)
How is the energy distributed in an equilibrium nonlinear chain at any
given instant of time, and how does this distribution depend
on the anharmonicity?  Can one talk about {\em spontaneous
energy localization} in thermal equilibrium, and, if so, what are the
mechanisms that lead to it? 2) How do local energy fluctuations in
such an equilibrium array relax in
a given oscillator? Are there circumstances in the equilibrium system
wherein a given oscillator remains at a high level of excitation for
a long time? 3)
Can local high-energy fluctuations move in some nondispersive fashion
along the array? Can an array in thermal equilibrium
transmit long-lived high-energy fluctuations
from one region of the array to another with little dispersion?

In our earlier work~\cite{Reigada1} we
showed that in harmonically coupled nonlinear chains
(``diagonal anharmonicity") in thermal
equilibrium, high-energy fluctuation mobility does {\em not} occur beyond
that which is observed in a harmonic chain.   Herein we show that the situation
might be quite different if there is ``nondiagonal anharmonicity," that
is, if the interoscillator interactions are anharmonic.

Our model consists of a one-dimensional array of $N$ unit-mass sites,
each connected by a potential $V(x_n-x_{n\pm 1})$
to its nearest neighbors that may be harmonic or anharmonic:
\begin{equation}
H=\sum_{n=1}^N \frac{p_n^2}{2} +\sum_{n=1}^N V(x_n-x_{n-1})~.
\label{hamiltonian}
\end{equation}
We assume periodic boundary conditions
and consider three prototype potentials:
\begin{alignat}{2}
\label{general1}
V_h(x)&=\frac{1}{2}kx^2 +\frac{1}{4} k' x^4 \qquad
  &{\rm hard},\\
\label{general2}
V_0(x)  & = \frac{1}{2} kx^2 \qquad &{\rm harmonic},\\
\label{general3}
V_s(x)   & = \frac{k}{k'}\left[ |x|-\frac{1}{k'}\ln (1+k'|x|)\right]
\qquad &{\rm soft} .
\end{alignat}
At small amplitudes the three
potentials are harmonic with the same force constant $k$. The
independent parameters $k$ and $k'$ allow
control of the harmonic component and the degree of anharmonicity of
the chain. Elsewhere~\cite{Sarmiento}
we have argued
that the overarching characteristic of anharmonic oscillators
is the dependence of
frequency on energy.  For a harmonic oscillator the
frequency is $\sqrt{k}$ independent of energy; for a hard
oscillator the frequency increases with energy, and for a soft
oscillator the frequency decreases with energy.

The set of coupled stochastic equations of motion for the masses
is that obtained from the
Hamiltonian, Eq.~(\ref{hamiltonian}), augmented by the usual
Langevin prescription for coupling a system to a heat bath at
temperature $T$:
\begin{equation}
\ddot{x}_n= -\frac{\partial}{\partial x_n}\left[ V(x_{n+1}-x_n) +
V(x_n-x_{n-1})\right]
- \gamma \dot{x}_n + \eta_n(t) ~,
\label{lang}
\end{equation}
where a dot represents a derivative with respect to time.
The $\eta_n(t)$ are mutually uncorrelated,
zero-centered, Gaussian,
$\delta$-correlated fluctuations that satisfy the fluctuation-dissipation
relation $\langle \eta_n(t) \eta_j(t') \rangle = 2 \gamma k_B T
\delta_{nj}\delta(t-t')$.
The numerical integration of the stochastic equations for all our
simulations is performed using the second order Heun's method
(which is equivalent to a second order Runge Kutta integration)
\cite{Gard,Toral}
with time step $\Delta t = 0.005$. In
each simulation the system is initially allowed to relax
for enough iterations to insure thermal equilibrium, after which we
take our ``measurements."

The equilibrium results to be presented here
complement our observations, presented
elsewhere, on the way in which these same chains propagate an energy
pulse~\cite{Sarmiento} as well as a sustained signal
applied at a particular site~\cite{Reigada3}.

\begin{figure}[htb!]
\begin{center}
\hspace{0.01in}
\epsfxsize = 4.in
\epsffile{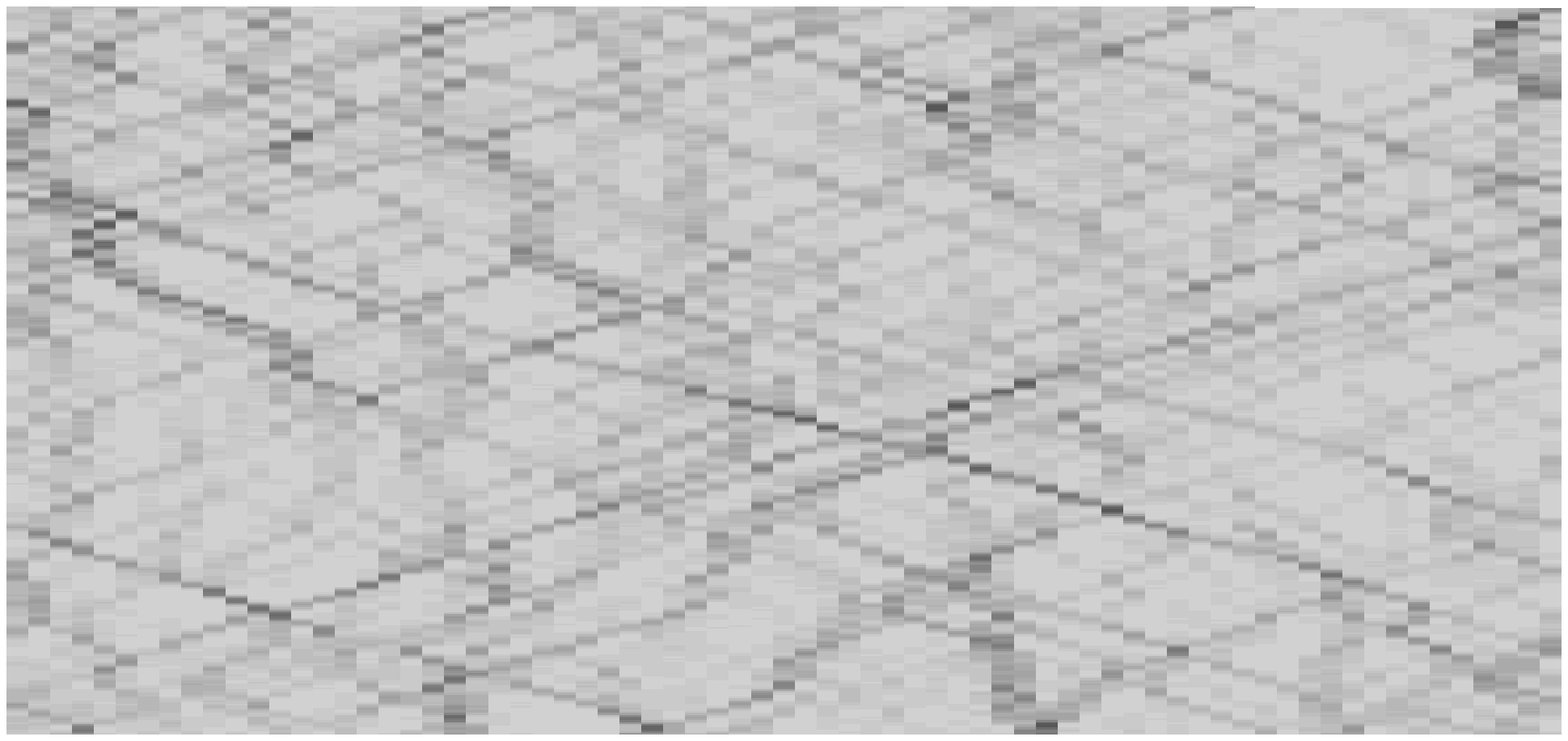}
\vspace{0.1in}
\epsfxsize = 4.in
\epsffile{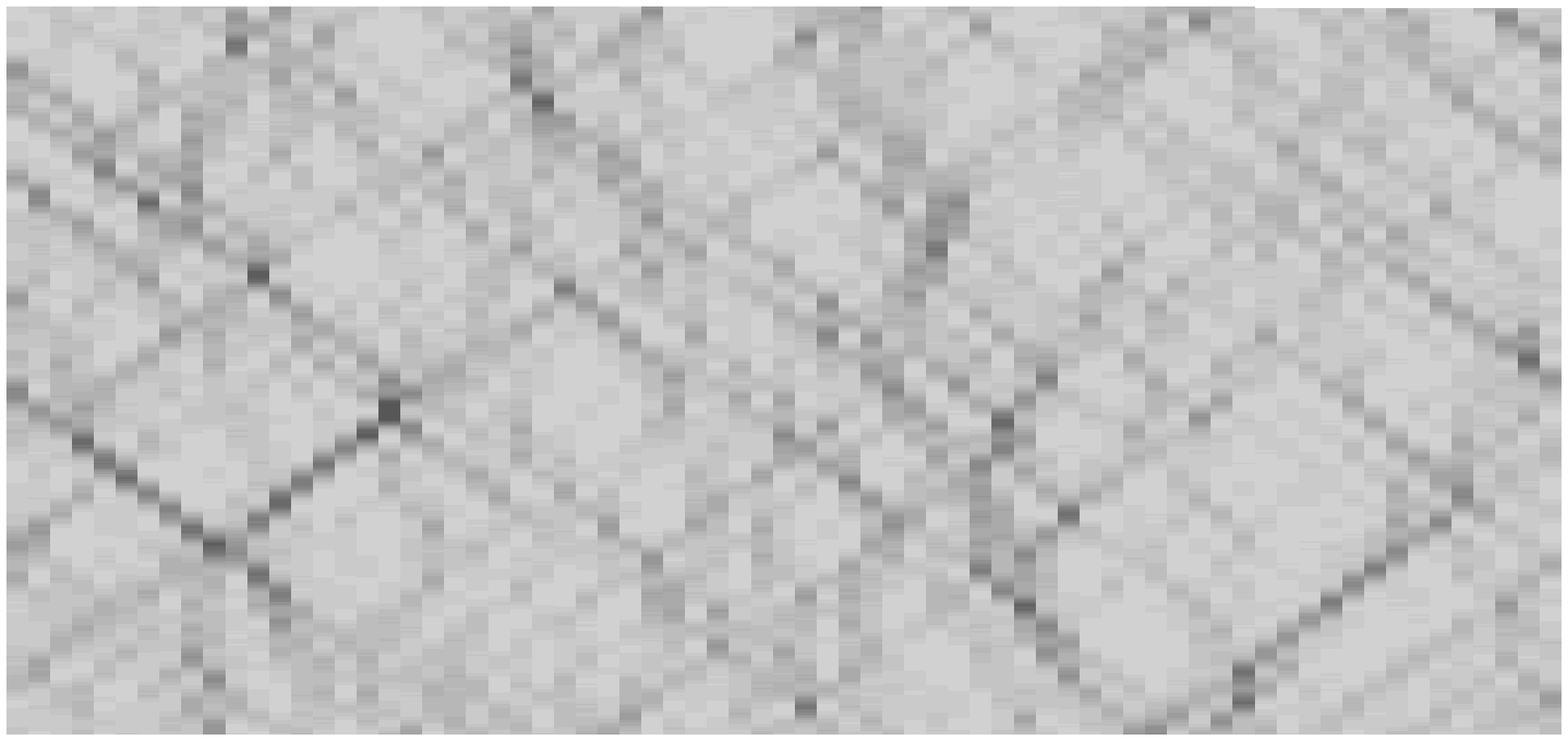}
\vspace{0.1in}
\epsfxsize = 4.in
\epsffile{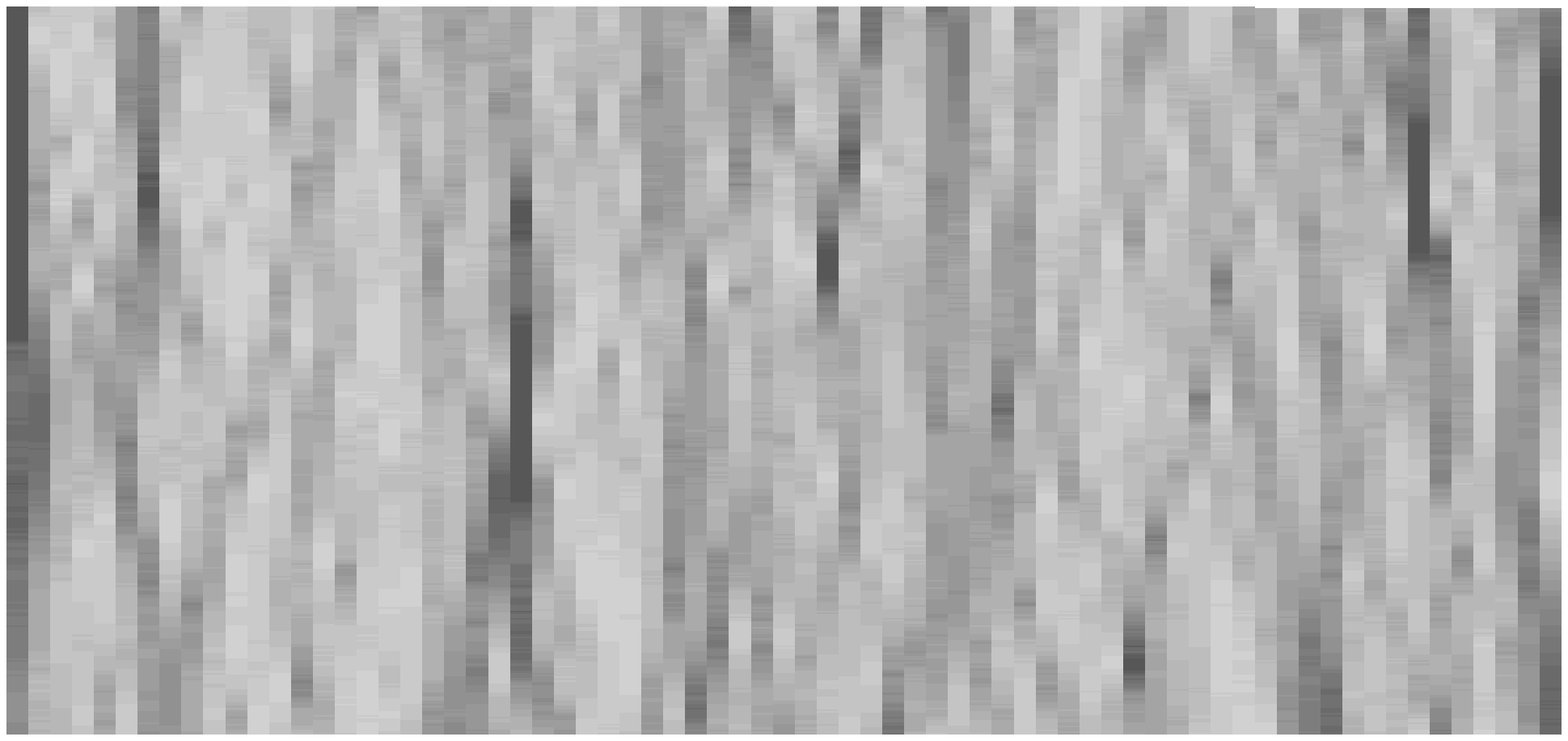}
\end{center}
\caption
{Energy (in grey scales) for thermalized chains of $71$
oscillators as a function of time.  The $x$-axis represents
the chain and time advances along the $y$-axis, with $t_{max}=160$.
The temperature is $k_BT=0.08$ and the dissipation parameter is
$\gamma=0.005$.
Top panel: hard chain with $k=0.1$, $k'=1$; middle panel: harmonic
chain with $k=0.1$; lower
panel: soft chain with $k=0.1$, $k'=5$.}
\label{landscape}
\end{figure}

A set of energy landscapes is shown in
Fig.~\ref{landscape}. Along the horizontal direction in
each panel lies a thermalized chain of oscillators; the
vertical upward progression shows the evolution of this
equilibrium system with time.  The gray scale represents the
energy, with darker shading reflecting more energetic regions.

Several noteworthy features are evident in the figure.
The energy fluctuations are greatest in the soft chain.  This
feature, seen earlier in chains with local anharmonic
potentials~\cite{Brown,Reigada1}, is a consequence of the effect
that we have called {\em entropic localization}.  In the soft
chain not only are the thermal fluctuations greater at a given
temperature, a result easily obtained from a
simple virial analysis, but the free energy is minimized by
a nonuniform distribution of energy that populates
regions of phase space where the density of states is high.
We have argued that
this localization mechanism is robust against temperature
increases -- indeed, it becomes more effective with increasing
temperature.  A second distinctive feature of the soft chain is
the persistence of the energy fluctuations: damping is not
particularly effective for soft chains. The only other mechanism
for removal of localized energy from a particular location is
along the chain.  This is clearly not an effective mechanism, a
result that is in agreement with our analysis of the propagation
of an externally applied pulse in the soft chain~\cite{Sarmiento}.
The speed of propagation
(in all chains) of a pulse of a given energy is essentially
proportional to the average frequency associated with that energy,
and in the soft chain this average frequency decreases with
increasing energy~\cite{Sarmiento}.  Although we do not see an obvious
connection between these excitations and solitons at zero temperature
(which are not entropic localization
mechanisms)~\cite{Willis,Takeno,Collins,Pnevmatikos,Scott,Cruzeiro}, 
there may be a closer connection with more 
generalized excitations such as breathers~\cite{Tsironis,Ebeling1,Bishop}.

In the hard chain (Fermi-Pasta-Ulam chain) the
total energy as well as the energy fluctuations are considerably
smaller but quite mobile with little dispersion, as can
easily be shown on the basis of
a straightforward statistical mechanical analysis.
In the hard chain the
average frequency increases with energy and therefore more
energetic pulses tend to travel more rapidly. We have also shown
that the dispersion of energy in a hard chain is
slow~\cite{Sarmiento} -- this is seen here in the integrity of
the spontaneously localized pulses over a much longer time than in
the harmonic chain.
\clearpage

\clearpage

An important question of course concerns the parameter regimes where
the differences illustrated in Fig.~\ref{landscape} are observed.
We have chosen potential parameters that insure clear distinctions in the
displacement amplitudes associated with the three potentials at the
chosen temperature.
The only restriction is that the temperature not be
``too low," that is, we avoid the region where all three
potentials are essentially harmonic.
We have chosen very low damping for the illustration.
The soft energy landscape is far less sensitive to damping than
the hard array.  An increase in damping would readily slow down
the motion of the high-energy fluctuations and would shorten their
lifetime.   Further, while the speed of the energy fluctuation pulses
is sensitive to the
potential parameters, their lifetime and dispersion properties are less so
(as long as one is in the highly anharmonic regime).  On the
other hand, the persistence of the fluctuations in the soft array is
quite sensitive to the harmonic contribution to the potential.
All else remaining fixed, the
landscapes remain qualitatively similar as temperature increases: the
fluctuations in the soft array become even stronger relative to the others,
and the pulse speeds in the hard array become even higher.

Suppose now that a bistable ``impurity" is embedded in each of
these chains, as illustrated in Fig.~\ref{drawings}.  When sufficient
energy reaches the impurity, a
transition may occur from one well to the other.
The statistical and dynamical properties of these transitions are
not obvious, and are explored in the next section.

\section{Kramers Problem and Statement of our Variant}
\label{kramers}

\subsection{Traditional Kramers Problem}
\label{traditional}

\begin{figure}[htb]
\begin{center}
\hspace{3.0in}
\epsfxsize = 3.3in
\epsffile{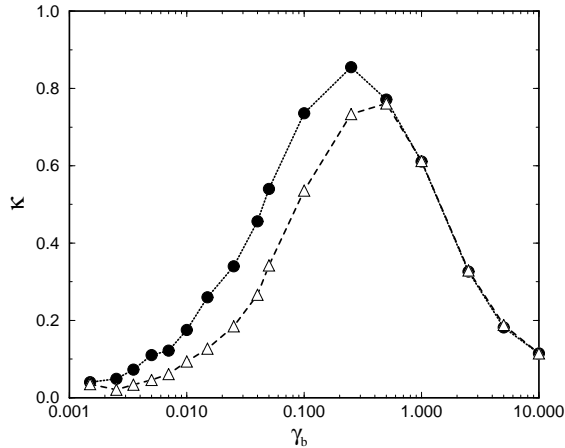}
\vspace{-0.5in}
\end{center}
\caption
{
Transmission coefficient $\kappa$ {\em vs} dissipation
parameter $\gamma_b$
for two temperatures
obtained from direct simulation of Eq.~(\ref{usualkramers}).
Solid circles: $k_BT = 0.025$; triangles:
$k_BT=0.05$.}
\label{kramersturnover}
\end{figure}

The original Markovian Kramers problem~\cite{Kramers} describes the
reaction coordinate $y$ evolving in the bistable potential
\begin{equation}
V_b(y)=\frac{V_0}{4}(y^2-1)^2
\label{bistable}
\end{equation}
according to the usual Langevin prescription
\begin{equation}
\ddot{y} = -\frac{dV_b(y)}{dy} -\gamma_b\dot{y} +\eta_b(t)
\label{usualkramers}
\end{equation}
where $\gamma_b$ is the dissipation parameter (the subscript for
``bistable" distinguishes this from the other dissipation
parameters) and $\eta_b(t)$ represents Gaussian, zero-centered,
$\delta$-correlated fluctuations that satisfy the
fluctuation-dissipation relation appropriate for temperature $T$,
$\langle \eta_b(t) \eta_b(t') \rangle = 2 \gamma_b k_B T \delta(t-t')$.
The rate coefficient $k_r$ for transitions from
one metastable well to the other is expressed as
\begin{equation}
k_r=\kappa \ k^{TST}~,
\end{equation}
where $k^{TST}$ is the rate obtained from transition state theory for
activated crossing, which in our units ($V_0=1$, frequencies $\sqrt{2}$
at the bottom of the two wells and unit frequency at the top of the
barrier) is
\begin{equation}
k^{TST}=\frac{\sqrt{2}}{\pi} \ e^{-1/4k_BT}~.
\end{equation}
This is the highest possible rate because it assumes no recrossings of
the barrier when the particle moves from one well to the other.  The
``transmission coefficient" $\kappa <1$ captures the effects of recrossings.
The dependence of $\kappa$ on
the various parameters of the problem has been the subject
of intense study over many
years~\cite{MelnikovHanggi,Melnikov1,Pollak1,Linkwitz}.
Its dependence on $\gamma_b$ and
temperature is exemplified in the simulations shown in
Fig.~\ref{kramersturnover}.  In particular, we note the occurrence
of a maximum:
as predicted by Kramers, the transmission coefficient at high
friction (diffusion-limited regime)
decays as $\gamma_b^{-1}$ (and is independent of temperature);  at
low friction (energy-limited regime) Kramers predicted that $\kappa$
is proportional to $\gamma_b/k_BT$.

An important generalization of the original Kramers problem, the so-called
Grote-Hynes problem~\cite{Grote}, reformulates
the model in terms of the generalized Langevin equation
\begin{equation}
\ddot{y} = -\frac{dV_b(y)}{dy} -\int_0^tdt' \Gamma(t-t')\dot{y}(t')+\eta_b(t)
\label{grotehynes}
\end{equation}
where $\Gamma(t-t')$ is a dissipative memory kernel and the
fluctuation-dissipation relation is now generalized to
$\left<\eta_b(t)\eta_b(t')\right>=k_BT\Gamma(t-t')$.  The dissipative
memory kernel reflects the dynamics of the thermal environment and is
characterized by its own time scales.
A frequent choice is an exponential,
but other forms that
have been used include a Gaussian and a decaying oscillatory
memory kernel.  The introduction of additional parameters of
course changes the behavior of the transmission coefficient.

The main point here is to call attention to the fact that
the transmission coefficient has the same value for two different values
of the dissipation parameter $\gamma_b$, and that therefore one can not
conclude whether the system is in one regime (diffusion-limited) or another
(energy-limited) simply from the value of the transition rate.
One needs to know the trend with changing dissipation parameter,
and one requires further information about the dynamics underlying a given
transition rate.  Not surprisingly, these dynamics turn out to be
entirely different in different regimes~\cite{Montg}.
The time dependence of the transmission coefficient
is a direct reflection of the
explicit trajectories of the particle as it transits from one well
to the other. A number of
investigators have looked at the time dependence of the
transmission coefficient in the
diffusion-limited~\cite{Borgis,Kohen,wekramers1,wekramers2,wekramers3}
and energy-limited~\cite{Borgis,wekramers1,wekramers2,wekramers3}
regimes, and also at the effect of different types of memory
kernels~\cite{wekramers2,wekramers3,Tucker}.

\begin{figure}[htb]
\begin{center}
\hspace{0.01in}
\epsfxsize = 4.in
\epsffile{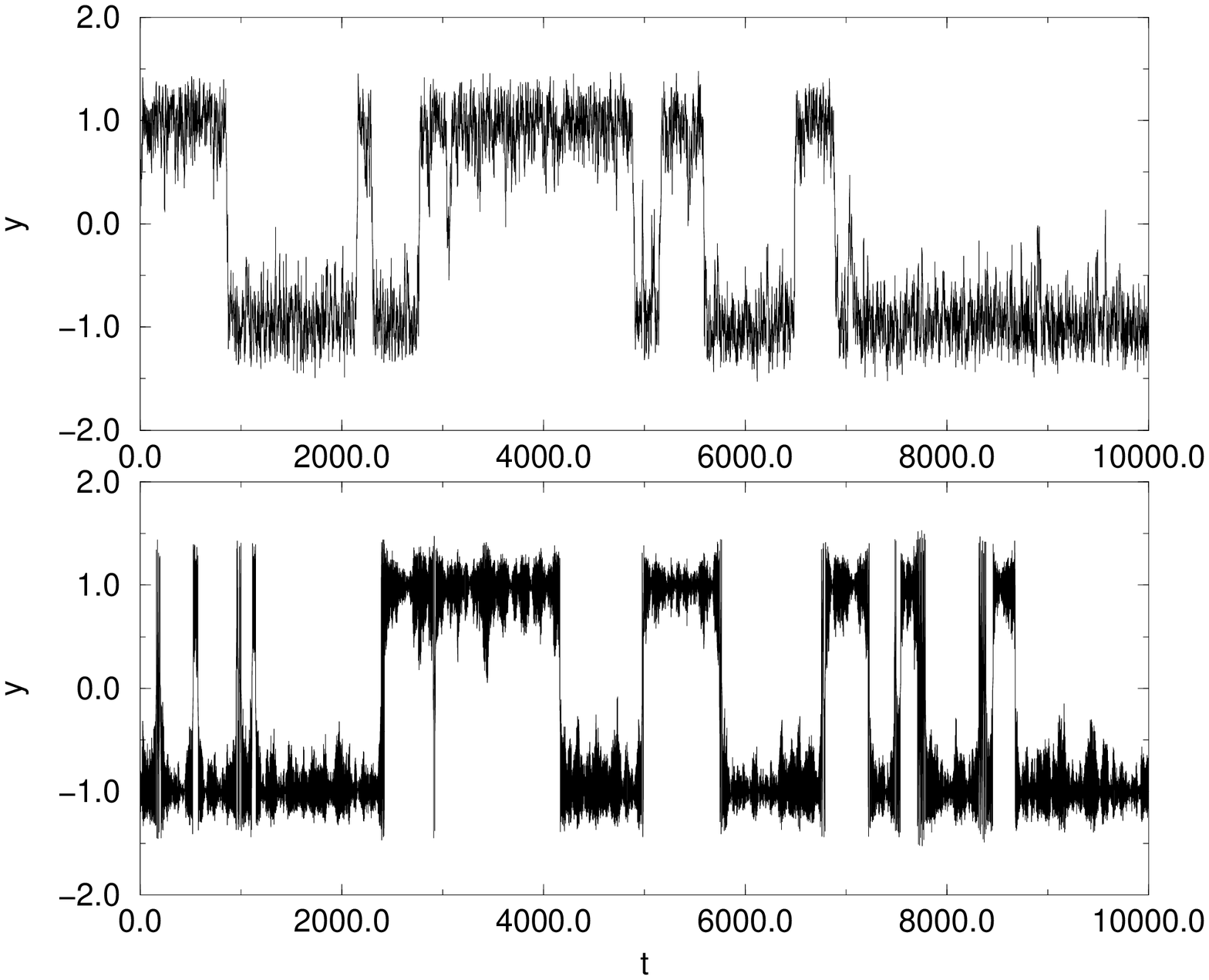}
\vspace{0.1in}
\epsfxsize = 4.in
\epsffile{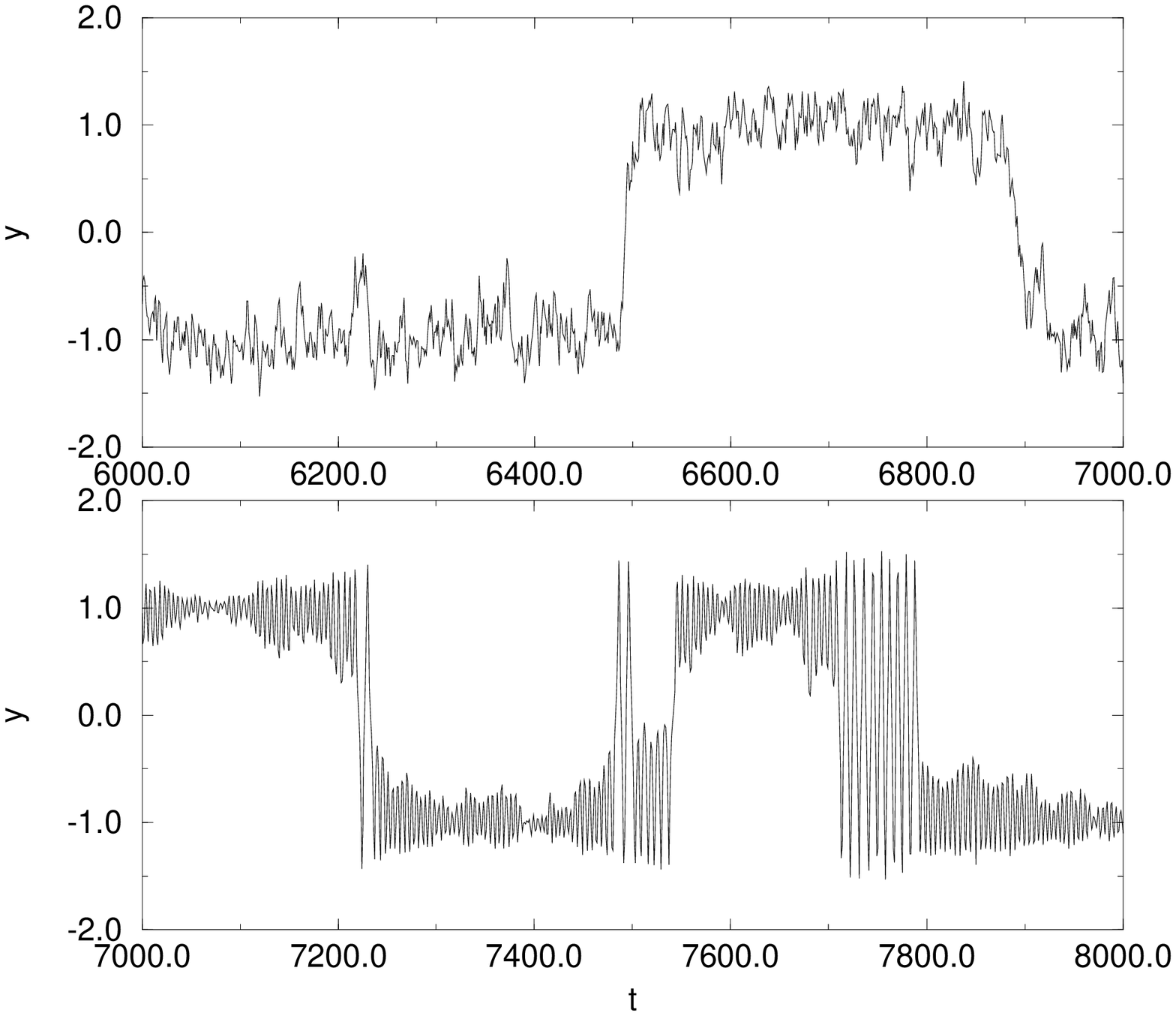}
\vspace{-0.3in}
\end{center}
\caption
{Trajectory of a bistable impurity described by the Langevin
equation, Eq.~(\ref{usualkramers}). The temperature is $k_BT=0.08$.
First panel: $\gamma_b=5.0$. Second panel: $\gamma_b=0.02$.  The
third (fourth) panel zooms in on a portion of the high-damping
(low-damping) trajectory.}
\label{kramerstrajectory}
\end{figure}

Of interest to us here are the different
dynamical behaviors in the diffusion-limited regime and
the energy-limited regime.
In Fig.~\ref{kramerstrajectory} we show two views of each of
two typical trajectories of
the reaction coordinate for the Markovian Kramers problem.  The
transmission coefficients associated with these two trajectories
are not too different (see below), but 
\clearpage
\noindent
they correspond to different
damping, placing them
on opposite sides of the turnover in the $\kappa$ vs
$\gamma_b$ curve.  The trajectory in the first panel is in the
diffusion-limited regime; the third panel shows an expanded
view of a portion of this trajectory. The particle
performs rather erratic motion within one well and once in a while
it surmounts the barrier.  When the particle surmounts
the barrier it does not spend much time in the barrier region before being
trapped again in one or the other well.  The crossing trajectories thus
tend to involve only one or a very small number of crossings/recrossings.
The trajectory in the second panel, a portion of which is
expanded in the fourth panel, is energy-limited. The particle
performs a fairly periodic motion within one well.  Barrier crossing events
tend to retain the particle in the barrier region for several recrossings;
a phase space analysis shows that the associated trajectories are
rather smooth oscillations from one side to the other of the potential
well above the barrier~\cite{wekramers1}.  Correlation functions
associated with these trajectories are presented and discussed
in Section~\ref{results}.

\subsection{Variant of the Kramers Problem}
\label{variant}

We would like to understand the way in which the very different thermal
landscapes described in Section~\ref{chains} affect the dynamics
of a reaction coordinate evolving in a
bistable ``impurity" embedded in these environments.  The connection of the
bistable impurity to the thermal environment occurs only through its
connection to the chain, that is, we set $\gamma_b=0$.

We need to specify how the bistable
system interacts with the chain.
We insert the impurity along the chain between sites $i$ and $i+1$
and connect it to each of these two
sites (see Fig.~\ref{drawings}). It is customary to
choose a simple interaction potential with a harmonic dependence
$V_{int}(x,y)\propto (x-y)^2$ for each
chain site connected to the impurity. Here
$y$ is the reaction coordinate and $x$ stands for the coordinate of
the chain site connected to the impurity.  However, this
interaction tends to destabilize the bistability in that
it causes the neighbors to pull the bistable particle {\em toward} the
barrier rather than toward its natural metastable states.
The interaction thus lowers the barrier of the bistable impurity.
Since we do not want to ``bias" the problem in this way, we have chosen
an interaction that instead tends to favor the already metastable
states:
\begin{equation}
V_{int}(x,y)=\frac{k_{int}}{2}\left( \frac{y^2-1}{2} -x\right)^2~.
\label{interaction}
\end{equation}
Near the bistable minima (which are shifted by the interaction)
the total potential for the reaction coordinate is still harmonic,
and near the maximum at $y=0$ it is still parabolic.
The barrier
height is modulated by the motion of the neighbors (somewhat reminiscent
of the barrier fluctuations in resonant activation problems).
At large values
of $y$ the interaction hardens the bistable potential.

The equations of evolution then have the following
contributions.  For a site in the chain not connected to the
bistable impurity we have, as before, Eq.~(\ref{lang}).
For the bistable impurity
\begin{equation}
\ddot{y} = -\frac{dV_b(y)}{dy} -\frac{\partial V_{int}(x_{i+1},y)}{\partial
y}  -\frac{\partial V_{int}(x_{i},y)}{\partial y} ~.
\end{equation}
For the site to the left of the bistable impurity
\begin{equation}
\ddot{x}_i=-\frac{\partial V(x_i-x_{i-1})}{\partial x_i}
   -\frac{\partial V_{int}(x_i,y)}{\partial x_i}-\gamma \dot{x}_i +\eta_i
(t)
\end{equation}
and similarly for $x_{i+1}$.
Comparisons are made for the same temperature, damping
coefficients, and interaction parameter
($k_{int}=0.1$ throughout this work) varying only the nature of the
chain.

\begin{figure}[htb]
\begin{center}
\epsfxsize = 6.in
\epsffile{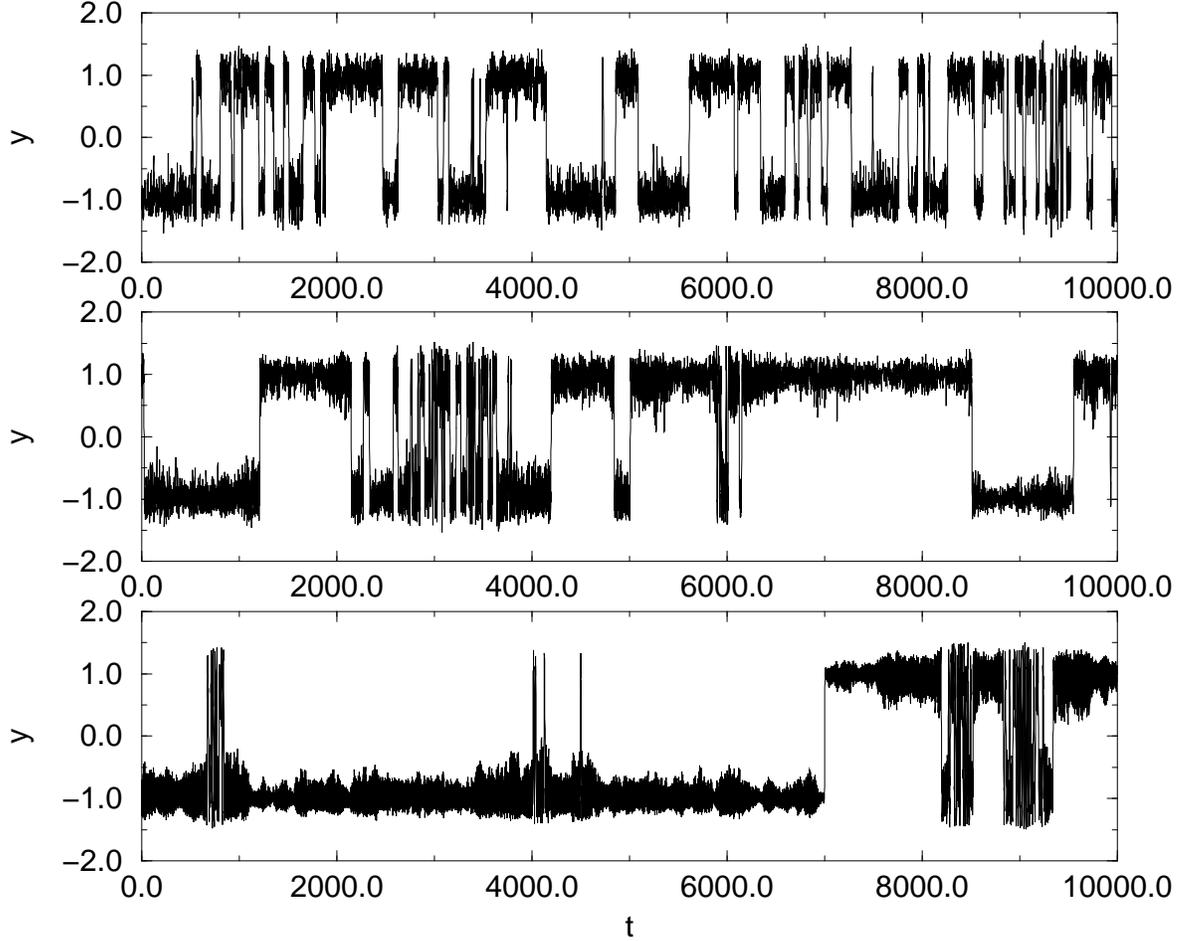}
\end{center}
\caption {Trajectory of a bistable impurity
embedded in a chain of $30$
oscillators with $k_BT =0.08$ and $\gamma = 0.005$.  Top panel: hard
chain with
$k=0.1$ and $k'=1$. Middle panel: harmonic chain with $k=0.1$. Bottom
panel: soft chain with $k=0.1$ and $k'=5$.}
\label{bistable005}
\end{figure}

Figure~\ref{bistable005} shows trajectories of the bistable
impurity embedded in each of the three chains. In the
hard chain the trajectory is rather similar to that of a Markovian
Kramers particle in
the diffusion-limited regime, while in the soft chain it is closer
to that of
the energy-limited regime.  This is a direct reflection of the behavior
seen in Fig.~\ref{landscape},
that is, of the fact that
in the hard chain independent thermal fluctuations created elsewhere
along the chain
have a good chance of reaching the bistable impurity, causing
erratic motion. An occasional large fluctuation causes a
transition over the barrier, usually unaccompanied by
recrossings: the same energy mobility that brings independent
fluctuations to the impurity also makes it easy for
the impurity to then lose a particular energy fluctuation back to
the chain.
In the soft chain, on the other hand, the particle performs fairly
periodic motion within one well.  Only fluctuations
in the sites immediately adjacent to the impurity can
excite the impurity; fluctuations originating elsewhere do not travel to
the impurity.  Strong fluctuations are therefore rarer but
more energetic and more persistent, so transition events occur
less often.  However, once such a fluctuation occurs it
tends to remain in that region for a long time; the reaction coordinate
therefore
recrosses the barrier a large number of times
until it eventually loses the
excess energy and is trapped again in one of the wells.

\begin{figure}[htb]
\begin{center}
\epsfxsize = 6.in
\epsffile{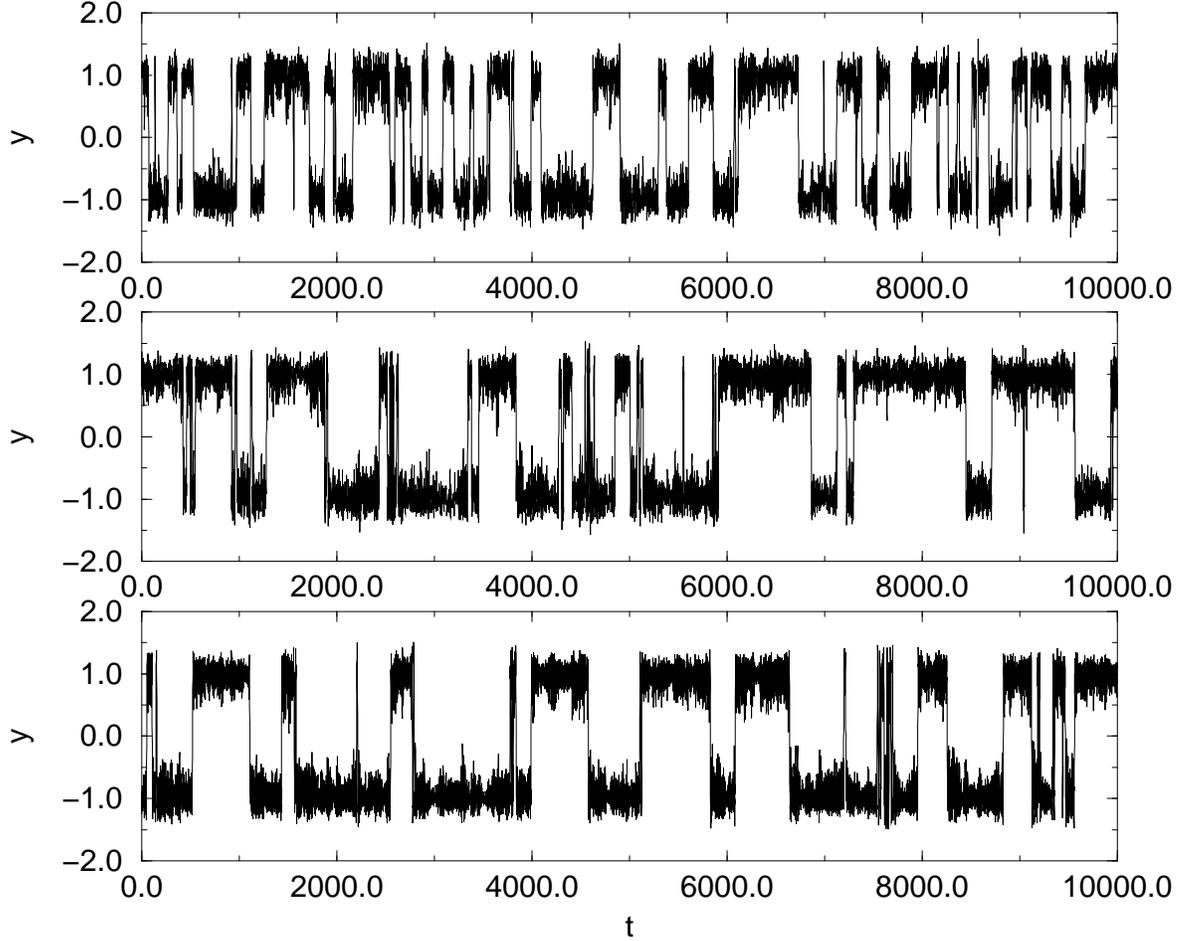}
\end{center}
\caption {Trajectory of a bistable impurity embedded in a chain
of $30$ oscillators.  All parameters are
the same as in Fig.~\ref{bistable005} except that the
dissipation parameter has been increased to $\gamma = 0.05$.}
\label{bistable05}
\end{figure}

A second set of trajectories associated with the same bistable
impurity in the same three chains at the same temperature but with
a (10-fold) higher dissipation parameter is
shown in Fig.~\ref{bistable05}.  Not surprisingly, the
trajectories are now more similar to one another, but nevertheless
there are still important and revealing differences that will
be made evident
in our discussion in the next section. Furthermore, a comparison
of the two sets will allow important observations concerning
the trends associated with increased damping.

In the next section we provide a quantitative characterization of
the differences in the trajectories and a comparison of these results with
those of the traditional Kramers problem.

\section{Results for Transition Rates}
\label{results}

A useful description of the bistable system
in different regimes is provided by the normalized correlation function
\begin{equation}
C(\tau)\equiv\frac{\left<
y(t+\tau)y(t)\right>}{\left<y^2(t)\right>}
\label{correlation}
\end{equation}
where the brackets indicate an average over $t$.  Since
$\left<y(t)\right>=0$, this correlation function decays to
zero. When the thermal
environment strongly and rapidly changes the particle momentum,
the trajectory is erratic and the correlation function decays
monotonically and exponentially.  The decay time is a measure
of the mean time between crossing events from one well to the other,
and its inverse can be identified with the transition rate $k_r$.
If on the other hand the effects of the thermal environment are weak,
then the trajectory is determined mainly by the deterministic potential and
remains correlated over much longer periods of time.

\begin{figure}[htb]
\begin{center}
\leavevmode
\epsfxsize = 3.0in
\epsffile{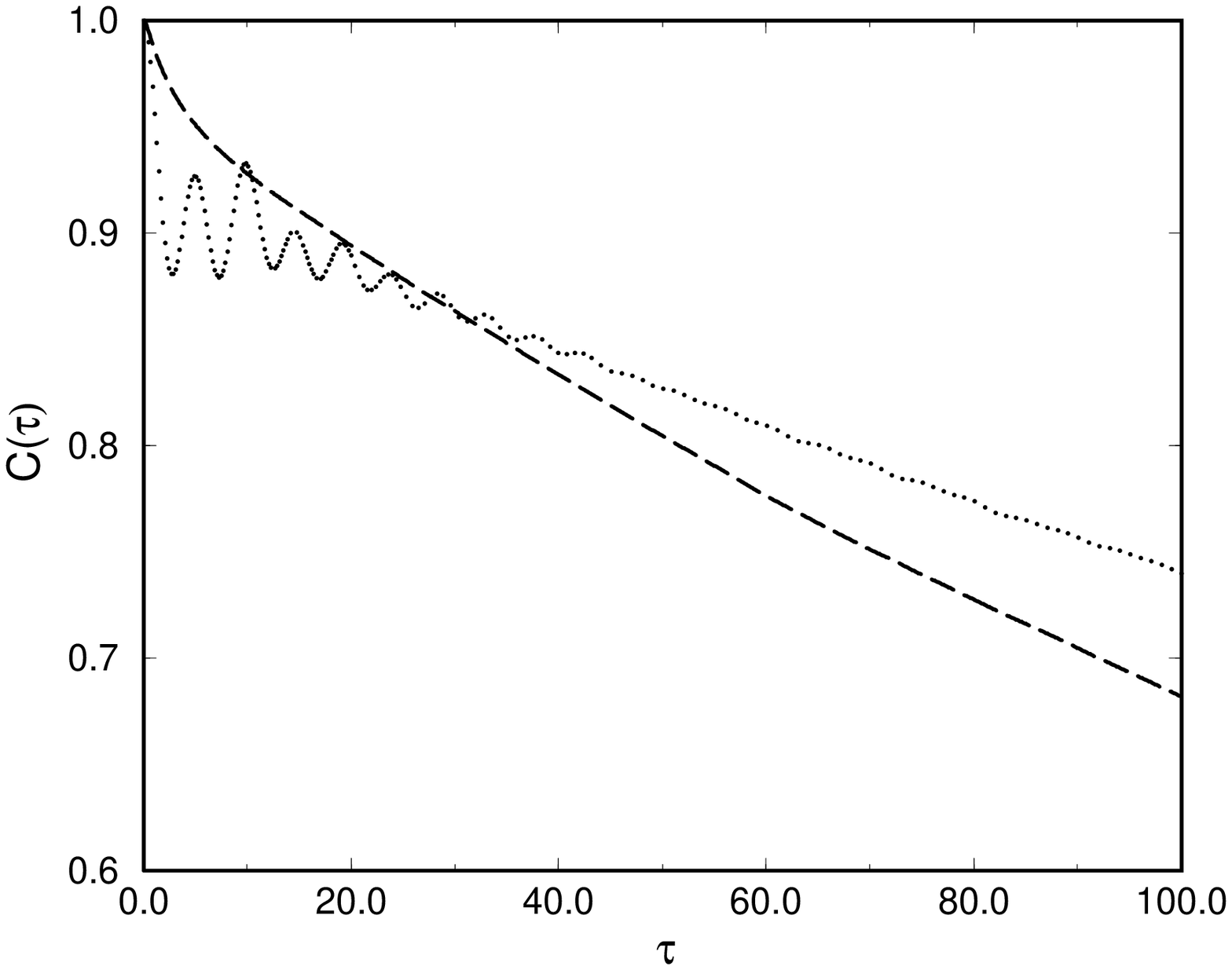}
\leavevmode
\epsfxsize = 3.0in
\epsffile{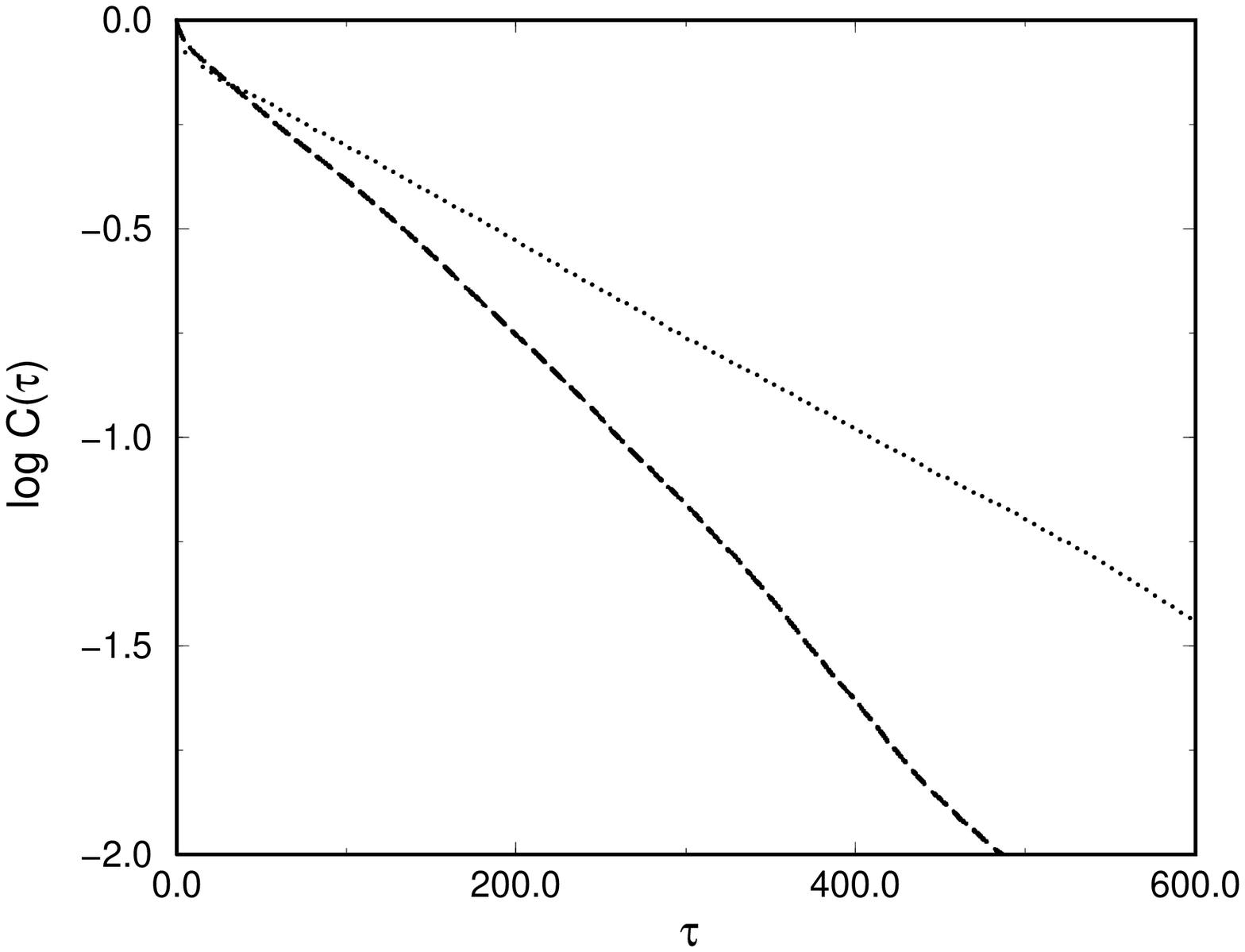}
\vspace{-0.2in}
\end{center}
\caption
{Correlation functions for the Markovian Kramers problem associated
with the trajectories of Fig.~\ref{kramerstrajectory}. Dashed curves:
$\gamma_b=5.0$, dotted curves: $\gamma_b=0.02$.
First panel: short-time behavior.
Second panel: correlation functions on a logarithmic time scale.}
\label{corrkramers}
\end{figure}

The correlation functions for the Markovian Kramers trajectories
of Fig.~\ref{kramerstrajectory} are shown in Fig.~\ref{corrkramers}.
In the high-dissipation regime the correlation function is monotonic
and decays exponentially over essentially all times.  This is a
reflection of the essentially random motion within each well and between
wells (the correlation functions for portions of the trajectory
entirely within one well are also monotonically decreasing, albeit not to
zero).  The slope of the high-$\gamma_b$ curve in the right panel
leads to a mean time between crossings of $\tau_c\approx 250$.

The oscillations in the low-$\gamma_b$ correlation function
reflect mainly
the systematic periodic motion of the particle within each
well, i. e., of the portions
of the trajectory that evolve for a long time near $y=1$ or near $y=-1$.
The period of these oscillations for
the parameters used here is $t_{bottom}=\sqrt{2}\pi$, and
this is very nearly the period of the oscillations in the figure.
Crossing events from one well to the other are mostly
separated by long times and are essentially independent (however, see
further discussion below).  Hence the logarithmic
rendition in the right panel gives a straight line.  Its slope leads to
a mean time between crossing events of $\tau_c\approx 450$.

\begin{figure}[htb]
\begin{center}
\leavevmode
\epsfxsize = 3.0in
\epsffile{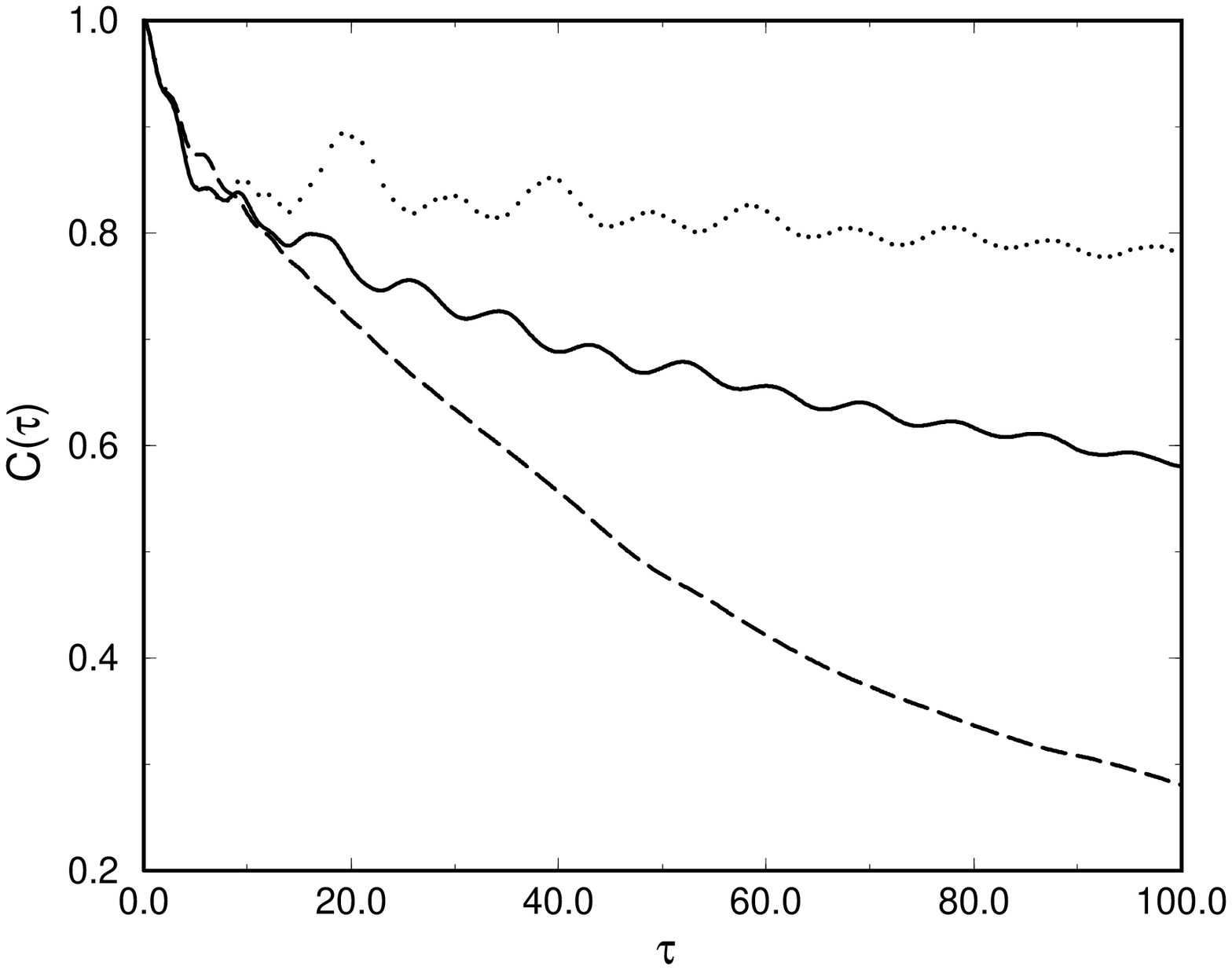}
\leavevmode
\epsfxsize = 3.0in
\epsffile{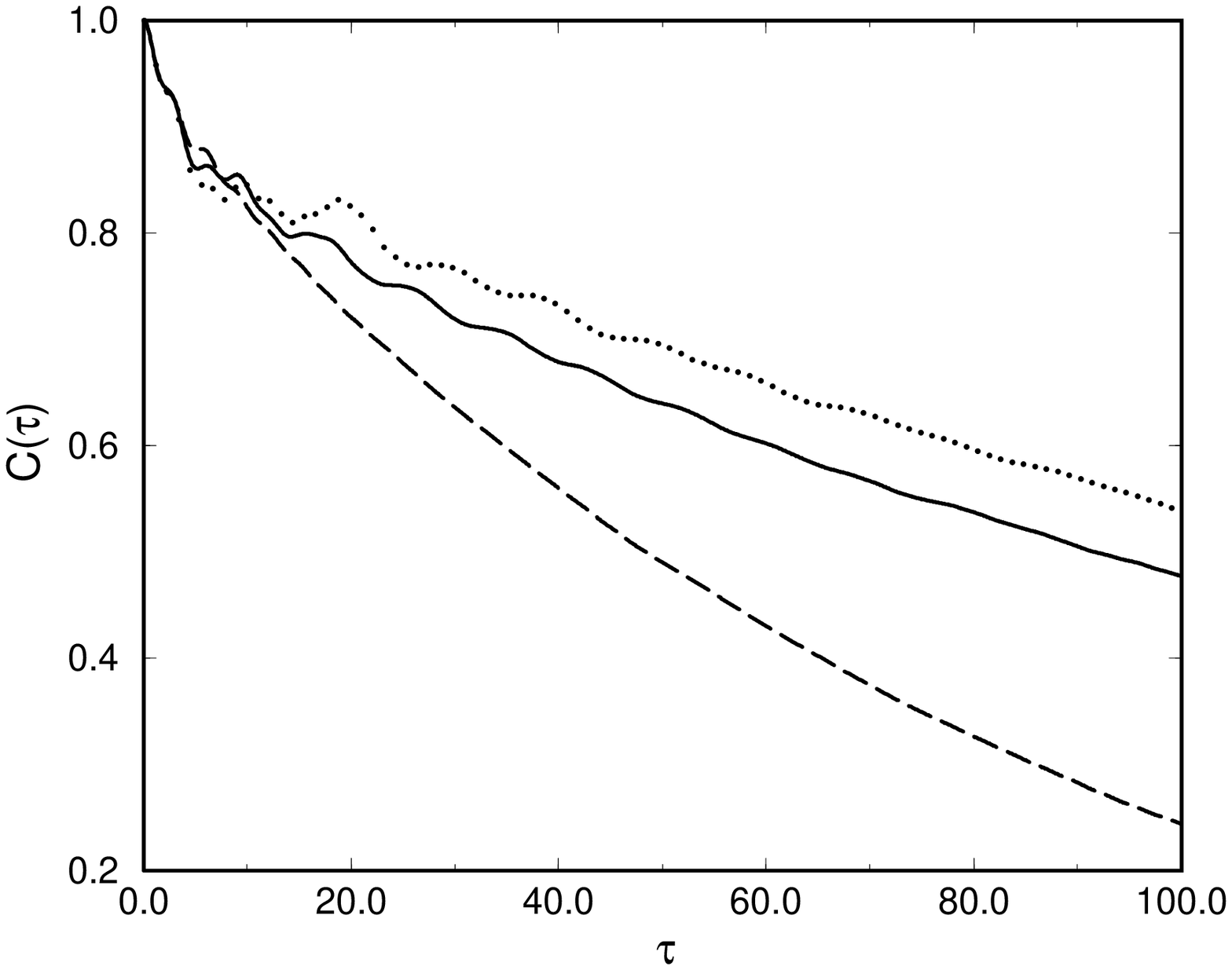}
\vspace{-0.2in}
\end{center}
\caption
{Correlation functions associated with the trajectories of
Figs.~\ref{bistable005} (first panel) and \ref{bistable05}
(second panel). Dashed curves: hard chain;
solid curves: harmonic chain; dotted curves: soft chain.}
\label{corrchains00505}
\end{figure}

The Kramers correlation functions serve as a point of reference
for an interpretation of the correlation functions associated with
our variant of the Kramers problem. These
are shown in a number of figures starting with Fig.~\ref{corrchains00505},
which
shows the correlations functions associated with the trajectories in
Figs.~\ref{bistable005} and \ref{bistable05}.
We stress that in each panel the $\gamma$ and $k_BT$ are the same
in all cases,
as is the coupling of the chain to the bistable system; only the nature
of the chain has changed.
The first panel shows 
a correlation function for the harmonic chain that is oscillatory
at early times, and quite similar to the energy-limited
Markovian Kramers case (see also the corresponding trajectories in
Figs.~\ref{bistable005} and \ref{kramerstrajectory}).  We conjecture that
the harmonic chain provides a thermal environment comparable to the
low-damping Markovian Kramers environment.  The correlation
function associated with
the hard chain is similar to the
behavior at higher damping in the Kramers case.
The correlation function associated with the
soft chain also decays in an oscillatory fashion, but in a more
complex way than in the energy-limited Markovian Kramers case.
{\em The alternation in the amplitudes
is a consequence of the presence of sustained bursts of energy that
cause a finite fraction of the trajectory to occur above the barrier,
leading to many correlated recrossing events}.
The particle oscillates above the barrier
for intervals much longer than in the
Markovian Kramers trajectory.  The typical oscillation period
above the barrier is
about twice as long as $t_{bottom}$ in our example (detailed
discussions of these times can be found in our earlier
work~\cite{wekramers1,wekramers2,wekramers3}).
This effect is already slightly visible in the low-$\gamma_b$
Kramers correlation function
in Fig.~\ref{corrkramers}, but
it is much stronger in the soft chain.
To reproduce this behavior in the Kramers model it is
necessary to consider the {\em generalized} Kramers model with a
memory friction: there is clearly
an additional memory effect in the soft chain that allows the energy
to remain trapped in the region of the bistable impurity for a long time.
This is in accord with the notion that transitions in the soft chain
are caused by local thermal fluctuations in the
nearest neighbors of the impurity. The impurity may periodically
exchange energy with these neighbors before the fluctuation eventually
dissipates away, and this causes repeated recrossings.  This in turn
leads to the conclusion that the memory friction in the Kramers model
needed to reproduce the soft chain environment would most likely be
oscillatory~\cite{wekramers3}.

The periods of oscillations in the harmonic and soft chains are somewhat
different from those of the Kramers curve in Fig.~\ref{corrkramers} and from
each other.  This is due to differences in the effective potentials.

The eventual decay of the correlation functions for all three chains is
exponential. We find for the times between transition events
(single or bursts as appropriate) $\tau_c\approx
101$ for the hard chain, $\tau_c\approx 453$ for the harmonic chain, and
$\tau_c\approx 1540$ for the soft chain.

A similar set of correlation functions associated with the
higher-damping-parameter trajectories of Fig.~\ref{bistable05}
is shown in the right panel of Fig.~\ref{corrchains00505}.
The dynamics of the bistable system in the hard chain with
increasing damping
does not change in character, whereas the oscillations in the harmonic
and soft chains become less pronounced as these systems move toward the
diffusion-limited regime.
First we note that all the curves become steeper, which
translates to a {\em shorter} mean time between crossing events and
therefore a {\em higher} transition rate for all
three chains.  The specific values we obtain
are $\tau_c\approx 72$ for the hard chain,
$\tau_c\approx 170$ for the harmonic chain, and $\tau_c\approx 215$
for the soft chain.  The decrease
in time between crossing events is most pronounced for the soft chain.
This is consistent with the notion that the soft chain is in the
energy-limited regime where small increases in effective damping cause the
greatest increases in the transition rate (see Fig.~\ref{kramersturnover}).
The harmonic chain lies
closer in this sense to the turnover region, and the hard chain even
closer yet.   The second point is that this apparent trend
for the hard chain indicates that it, too, lies on the
low-damping side
of the turnover in spite of the diffusion-limited aspects of its
dynamics.
This is the reason for the very small oscillations visible at the earliest
times in the hard chain correlation functions.  It is
apparent that
neither the trajectory itself nor even the shape of the correlation
function at one value of the damping provides unequivocal
information to determine which side of the turnover regime one
is on; it is necessary to investigate the trend.

It is interesting to
investigate whether our variants of the thermal environment
can actually be ``pushed" across the
turnover point by increasing the damping on the chain.  For this purpose
we present a series of correlation functions
for each of the chains for different values of the damping parameter.
The other parameters, including the temperature, remain fixed
and equal to the values given earlier.

\begin{figure}[htb]
\begin{center}
\leavevmode
\epsfxsize = 3.0in
\epsffile{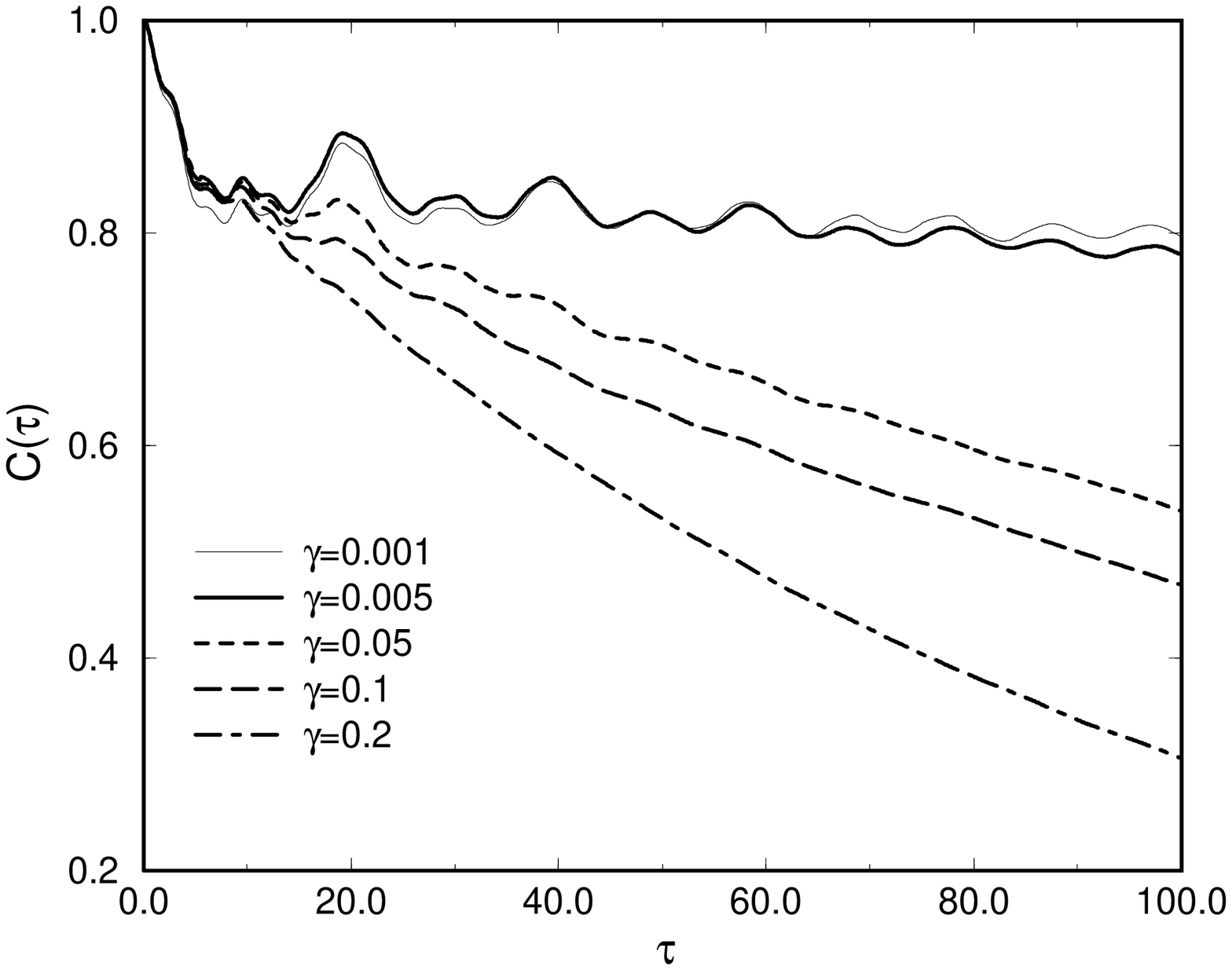}
\leavevmode
\epsfxsize = 3.0in
\epsffile{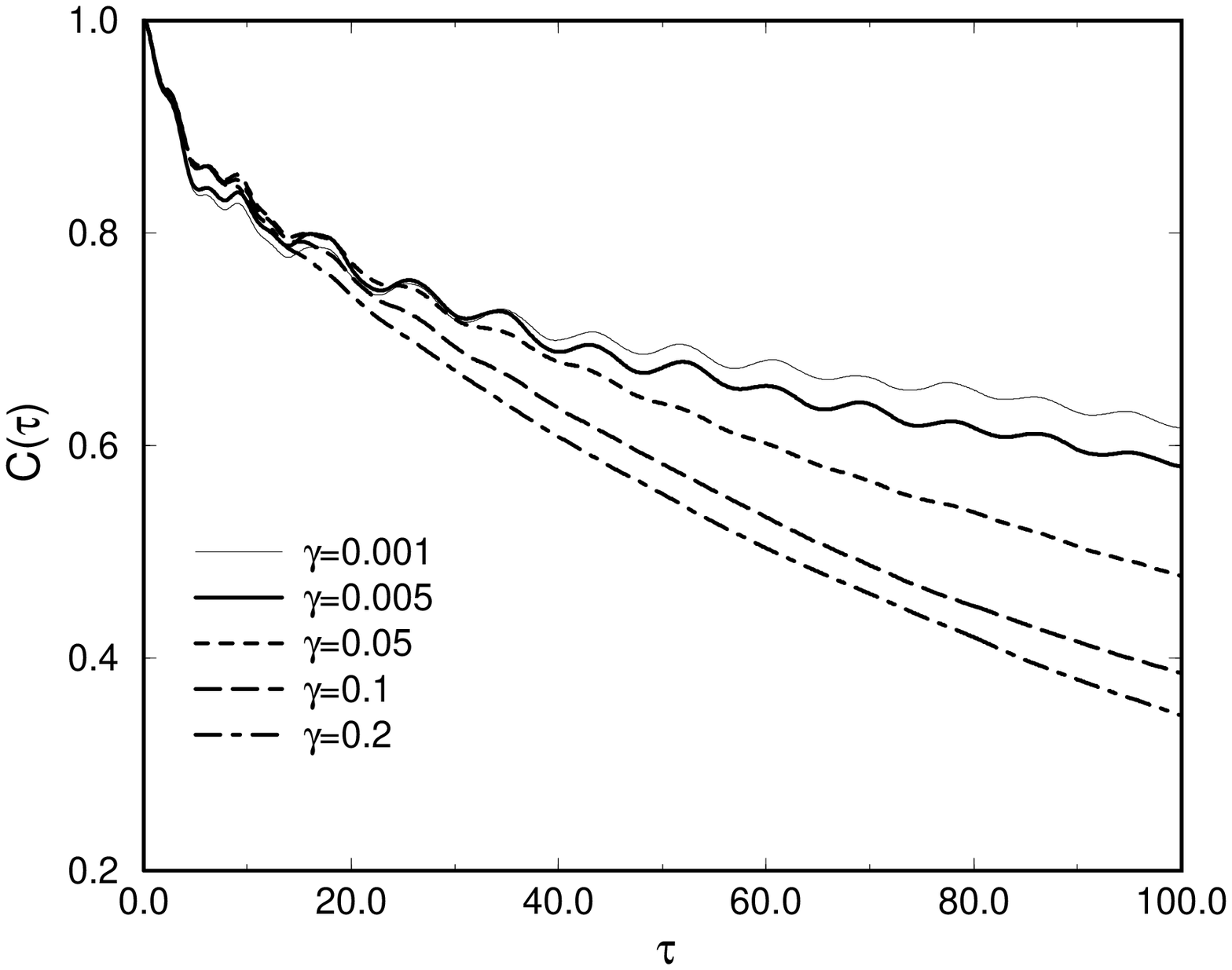}
\vspace{-0.2in}
\end{center}
\caption
{Correlation functions for the bistable impurity in the
soft chain (first panel) and harmonic chain (second panel)
or various values of the dissipation parameter.}
\label{corsoftharm}
\end{figure}

\begin{figure}[htb]
\begin{center}
\leavevmode
\epsfxsize = 3.0in
\epsffile{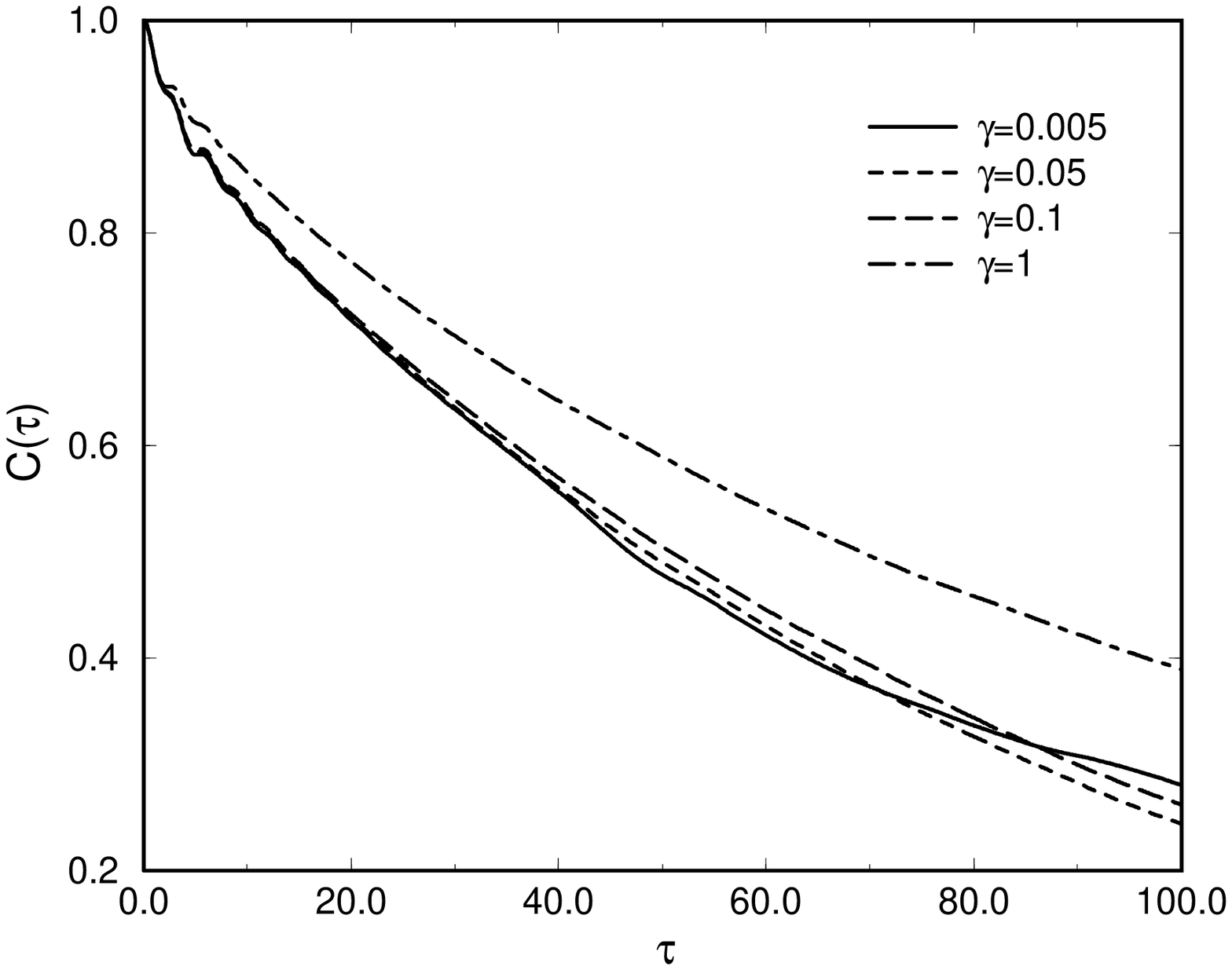}
\leavevmode
\epsfxsize = 3.0in
\epsffile{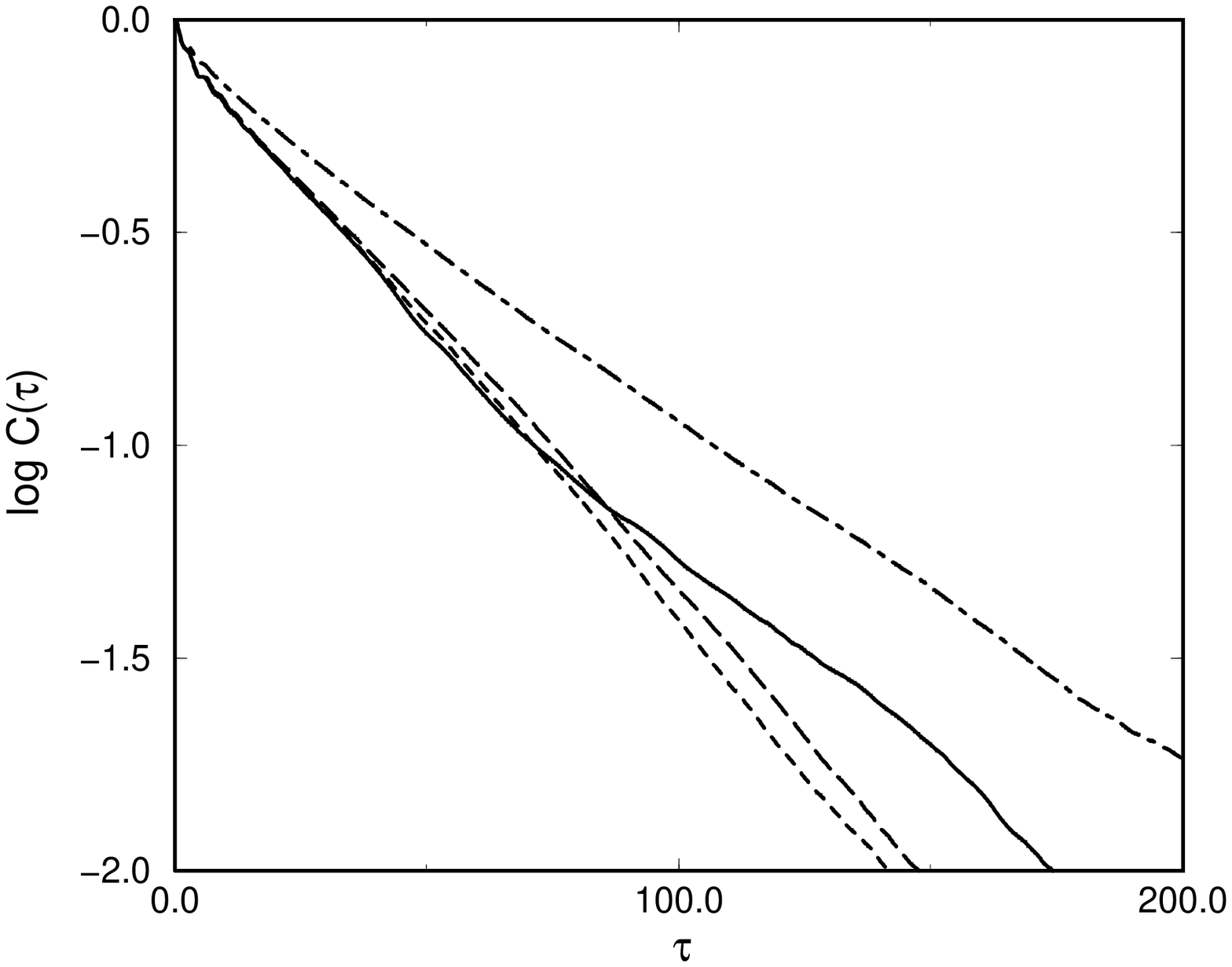}
\vspace{-0.2in}
\end{center}
\caption
{Correlation functions for the bistable impurity in the
hard chain for various values of the dissipation parameter.
First panel: short-time behavior.
Second panel: correlation function on a logarithmic scale.}
\label{corhard}
\end{figure}

Figure~\ref{corsoftharm} shows results for the soft and harmonic
chains.  With increasing
damping the early-time oscillations in the correlation function in
the soft chain first
lose some of the ``alternation" feature typical of a long oscillatory
dissipative memory kernel and eventually the correlation function
loses its oscillatory character
altogether.  The
crossing rate continues to increase as damping increases, so throughout
this series one is still on the low damping side of the
turnover. The right panel of
Fig.~\ref{corsoftharm} shows the correlation functions for the impurity
in the harmonic chain.  The trend is similar to that of the soft chain
but, in all respects, indicative of the fact that the harmonic
environment is closer to the turnover region than the soft environment.
Thus the oscillations disappear sooner, and the increase in the transition
rate with increasing damping is smaller.

Perhaps the most interesting features are seen in Fig.~\ref{corhard}.
Here we clearly see the very small short-time oscillations, which
disappear as damping increases. The transition
rate is quite insensitive to damping in the range $\gamma=0.005$ --
$0.1$ shown in the figure (the line for $\gamma=0.2$, not
explicitly shown, also falls in the same regime).
The turnover value must therefore be within this range.  To ascertain
if this is so, we also exhibit the correlation function for a considerably
larger value of the dissipation parameter, $\gamma=1.0$.
The slower decay for $\gamma=1$ is clear in both panels.

\section{Conclusions}
\label{conclusions}

There has been a dearth of information on
the effects on the activation process of {\em nonlinearities} in
the environment. We have taken an approach here
that goes part way, much in the tradition of modeling efforts for a
variety of
systems interacting with a complex environment: the ``immediate
surroundings" of the reaction coordinate are described microscopically,
while the interaction of this immediate environment with other degrees of
freedom is handled phenomenologically.

We find that the dynamics of the activation process in some parameter
regimes are profoundly affected by the nature of the chain.
If the
damping parameter connecting the chain to the heat bath
is sufficiently low, a soft
chain provides an environment very similar to that of the
Grote Hynes model with an oscillatory memory kernel in the
energy-limited regime~\cite{wekramers3}, while a hard chain
provides an environment
akin to that of the Kramers model in the diffusion-limited
regime~\cite{wekramers1}.
This in turn means that in such parameter regimes {\em the hard chain
is a more effective mediator of the activation process
than is the soft chain}.

A number of interesting questions concerning these
systems are currently under investigation.  One concerns the
influence of boundary conditions on the behavior that we have
described~\cite{Reigada2}.
A second problem concerns the effect on the reaction
coordinate of a pulse or a sustained signal applied somewhere else
along the chain.  We have showed that the
propagation of such a pulse or signal is strongly
affected by the nature of the chain~\cite{Sarmiento,Reigada3},
and we expect these
differences in turn to affect the response of a bistable impurity
to these excitations.  Such models are interesting in the context
of physical or biophysical situations wherein energy is released
at some location (provided perhaps by a chemical reaction or an
absorption process at that location), which must then move to
another location (that of the bistable impurity) to effect some
further chemical process (represented by the activation process).
The usual linear chain models are plagued by the excessive
dispersion that would make such transmission inefficient.
Nonlinearities in the environment may provide the necessary
mobility with little attendant dispersion, thus greatly increasing
the efficiency of such a process.

\section*{Acknowledgments}
R. R. gratefully acknowledges the support of this
research by the Ministerio de Educaci\'{o}n y Cultura through
Postdoctoral Grant No.\ PF-98-46573147.  A. S. acknowledges sabbatical
support from DGAPA-UNAM.
This work was supported in part by the Engineering Research Program of the
Office of Basic Energy Sciences at the U.S. Department of Energy under
Grant No. DE-FG03-86ER13606, and in part by the Comisi\'on
\nobreak{Interministerial} de Ciencia y Tecnolog\'{\i}a (Spain)
Project No.\ DGICYT PB96-0241.

\clearpage
\setcounter{figure}{0}

\begin{figure}
\caption
{Schematic of a bistable impurity (``reaction coordinate" RC) connected
to a chain that interacts with a heat bath at temperature $T$.  The chain
masses are indicated by rhombuses. Their interactions (shown schematically
as springs) may be harmonic or anharmonic.  Each mass in the chain is
subject to thermal fluctuations denoted by $\eta$ and the usual
accompanying dissipation. The bistable impurity interacts only with the
chain, which thus provides its thermal environment.
The bistable system is inserted in the chain (``disruptive impurity");
its detailed interaction with the chain is discussed in the text.}
\end{figure}

\begin{figure}
\caption
{Energy (in grey scales) for thermalized chains of $71$
oscillators as a function of time.  The $x$-axis represents
the chain and time advances along the $y$-axis, with $t_{max}=160$.
The temperature is $k_BT=0.08$ and the dissipation parameter is
$\gamma=0.005$.
Top panel: hard chain with $k=0.1$, $k'=1$; middle panel: harmonic
chain with $k=0.1$; lower
panel: soft chain with $k=0.1$, $k'=5$.}
\end{figure}

\begin{figure}
\caption
{
Transmission coefficient $\kappa$ {\em vs} dissipation
parameter $\gamma_b$
for two temperatures
obtained from direct simulation of Eq.~(\ref{usualkramers}).
Solid circles: $k_BT = 0.025$; triangles:
$k_BT=0.05$.}
\end{figure}

\begin{figure}
\caption
{Trajectory of a bistable impurity described by the Langevin
equation, Eq.~(\ref{usualkramers}). The temperature is $k_BT=0.08$.
First panel: $\gamma_b=5.0$. Second panel: $\gamma_b=0.02$.  The
third (fourth) panel zooms in on a portion of the high-damping
(low-damping) trajectory.}
\end{figure}

\begin{figure}
\caption {Trajectory of a bistable impurity
embedded in a chain of $30$
oscillators with $k_BT =0.08$ and $\gamma = 0.005$.  Top panel: hard
chain with
$k=0.1$ and $k'=1$. Middle panel: harmonic chain with $k=0.1$. Bottom
panel: soft chain with $k=0.1$ and $k'=5$.}
\end{figure}

\begin{figure}
\caption {Trajectory of a bistable impurity embedded in a chain
of $30$ oscillators.  All parameters are
the same as in Fig.~\ref{bistable005} except that the
dissipation parameter has been increased to $\gamma = 0.05$.}
\end{figure}

\begin{figure}
\caption
{Correlation functions for the Markovian Kramers problem associated
with the trajectories of Fig.~\ref{kramerstrajectory}. Dashed curves:
$\gamma_b=5.0$, dotted curves: $\gamma_b=0.02$.
First panel: short-time behavior.
Second panel: correlation functions on a logarithmic time scale.}
\end{figure}

\begin{figure}
\caption
{Correlation functions associated with the trajectories of
Figs.~\ref{bistable005} (first panel) and \ref{bistable05}
(second panel). Dashed curves: hard chain;
solid curves: harmonic chain; dotted curves: soft chain.}
\end{figure}

\begin{figure}
\caption
{Correlation functions for the bistable impurity in the
soft chain (first panel) and harmonic chain (second panel)
or various values of the dissipation parameter.}
\end{figure}

\begin{figure}
\caption
{Correlation functions for the bistable impurity in the
hard chain for various values of the dissipation parameter.
First panel: short-time behavior.
Second panel: correlation function on a logarithmic scale.}
\end{figure}

\end{spacing}
\end{document}